\documentstyle[12pt]{article}

\newcommand{\AmS}{{\protect\the\textfont2
  A\kern-.1667em\lower.5ex\hbox{M}\kern-.125emS}}
\newcommand{\bff}[1]{{\mbox{\boldmath $#1$}}}
\begin{document}
\title{\bf Relativistic mean-field description
       of the dynamics of giant resonances}
\author{\\
D. Vretenar $^{\rm a}$, P. Ring $^{\rm b}$, 
	G.A. Lalazissis $^{\rm b}$, and N.Paar $^{\rm a}$\\[2.0ex] 
$^{\rm a}$ Physics Department, Faculty of Science,\\
	University of Zagreb, Croatia\\[2.0ex]
$^{\rm b}$ Physik-Department der Technischen
	Universit\"at M\"unchen,\\
	D-85748 Garching, Germany\\}
\maketitle

\begin{abstract}
The relativistic mean-field theory provides a framework in which
the nuclear many-body problem is described as a self-consistent
system of nucleons and mesons.
In the mean-field approximation, the self-consistent
time evolution of the nuclear system
describes the dynamics of collective motion:
nuclear compressibility from monopole resonances,
regular and chaotic dynamics of isoscalar and isovector
collective vibrations.
\end{abstract}

\section{The relativistic mean field model}

Relativistic mean-field (RMF) models have been successfully applied in
calculations of nuclear matter and properties of finite nuclei throughout
the periodic table. In the self-consistent mean-field
approximation, detailed calculations have been performed for a variety
of nuclear structure phenomena~\cite{Rin.96}. In this
work we present applications of RMF to the
dynamics of collective vibrations in spherical nuclei. 
In relativistic quantum hadrodynamics~\cite{SW.97},
the nucleus is described as a system of Dirac nucleons which
interact through the
exchange of virtual mesons and photons.  The Lagrangian
density of the model is
\begin{eqnarray}
{\cal L}&=&\bar\psi\left(i\gamma\cdot\partial-m\right)\psi
~+~\frac{1}{2}(\partial\sigma)^2-U(\sigma )
\nonumber\\
&&-~\frac{1}{4}\Omega_{\mu\nu}\Omega^{\mu\nu}
+\frac{1}{2}m^2_\omega\omega^2
~-~\frac{1}{4}{\vec{\rm R}}_{\mu\nu}{\vec{\rm R}}^{\mu\nu}
+\frac{1}{2}m^2_\rho\vec\rho^{\,2}
~-~\frac{1}{4}{\rm F}_{\mu\nu}{\rm F}^{\mu\nu}
\nonumber\\
&&-~g_\sigma\bar\psi\sigma\psi~-~
g_\omega\bar\psi\gamma\cdot\omega\psi~-~
g_\rho\bar\psi\gamma\cdot\vec\rho\vec\tau\psi~-~
e\bar\psi\gamma\cdot A \frac{(1-\tau_3)}{2}\psi\;.
\label{lagrangian}
\end{eqnarray}
The Dirac spinor $\psi$ denotes the nucleon with mass $m$.
$m_\sigma$, $m_\omega$, and $m_\rho$ are the masses of the
$\sigma$-meson, the $\omega$-meson, and the $\rho$-meson,
and $g_\sigma$, $g_\omega$, and $g_\rho$ are the
corresponding coupling constants for the mesons to the
nucleon. $U(\sigma)$ denotes the nonlinear $\sigma$
self-interaction,
and $\Omega^{\mu\nu}$, $\vec R^{\mu\nu}$, and $F^{\mu\nu}$
are field tensors~\cite{Rin.96,SW.97}.
The coupled equations of motion are derived from the Lagrangian
density (\ref{lagrangian}).
The Dirac equation for the nucleons:
\begin{eqnarray}
i\partial_t\psi_i&=&\left[ \bff\alpha
\left(-i\bff\nabla-g_\omega\bff\omega-
g_\rho\vec\tau\vec{\bff\rho}
-e\frac{(1-\tau_3)}{2}{\bff A}\right)
+\beta(m+g_\sigma \sigma)\right.\nonumber\\ 
&&\left. +g_\omega \omega_0+g_\rho\vec\tau\vec\rho_0
+e\frac{(1-\tau_3)}{2} A_0
\right]\psi_i
\label{dirac}
\end{eqnarray}
and the Klein-Gordon equations for the mesons:
\begin{eqnarray}
\left(\partial_t^2-\Delta+m^2_\sigma\right)\sigma&=&
-g_\sigma\rho_s-g_2 \sigma^2-g_3 \sigma^3\\
\left(\partial_t^2-\Delta+m^2_\omega\right)\omega_\mu&=&
~g_\omega j_\mu\\
\left(\partial_t^2-\Delta+m^2_\rho\right)\vec\rho_\mu&=&
~g_\rho \vec j_\mu\\
\left(\partial_t^2-\Delta\right)A_\mu&=&
~e j_\mu^{\rm em}.
\label{KGeq4}
\end{eqnarray}
In the relativistic mean-field approximation, the nucleons
described by single-particle spinors $\psi_i$ are
assumed to form the A-particle Slater determinant $|\Phi\rangle$,
and to move independently in the classical meson fields.
The sources of the fields, i.e.
densities and currents, are calculated in the {\it no-sea}
approximation~\cite{VBR.95}:\ the scalar density:
$\rho_{\rm s}~=~\sum_{i=1}^A \bar\psi_i\psi_i$,
the isoscalar baryon current:
$j^\mu~=~\sum_{i=1}^A \bar\psi_i\gamma^\mu\psi_i$,
the isovector baryon current:
$\vec j^{\,\mu}~=~\sum_{i=1}^A \bar\psi_i\gamma^\mu \vec \tau\psi_i$,
the electromagnetic current for the photon-field:
$j^\mu_{\rm em}~=~\sum_{i=1}^A
\bar\psi_i\gamma^\mu\frac{1-\tau_3}{2}\psi_i$.
The summation is over all occupied states in the
Slater determinant $|\Phi\rangle$. Negative-energy states
do not contribute to the densities in the {\it no-sea}
approximation of the stationary solutions.  
It is assumed that nucleon single-particle
states do not mix isospin. 

The ground state of a nucleus is described by
the stationary self-consistent solution of the
coupled system of equations
(\ref{dirac})--(\ref{KGeq4}),
for a given number of nucleons
and a set of coupling constants and masses.
The solution for the ground state specifies part of the initial
conditions for the time-dependent problem.
The dynamics of nuclear collective motion is analyzed in the framework
of time-dependent relativistic mean-field model,
which represents a relativistic generalization of the
time-dependent Hartree-Fock approach. 
For a given set of initial conditions, i.e. initial values
for the densities and currents, nuclear dynamics is
described by the simultaneous evolution of $A$ single-
particle Dirac spinors in the time-dependent mean fields.
The equations (\ref{dirac})--(\ref{KGeq4})
are equivalent to the equation of motion for the
one-body density operator $\hat \rho = \hat \rho(t)$
\begin{equation}
i \hbar {\partial \over {\partial t}} \hat \rho =
\left[ h_D, \hat \rho \right] ,
\label{eom}
\end{equation}
with an initial condition for $\hat \rho$
\begin{equation}
\hat \rho(t_{in}) = \hat \rho_{in} .
\end{equation}
$h_D$ is the single-nucleon Dirac hamiltonian defined
in Eq.~(\ref{dirac}). Starting from the self-consistent solution
that describes the ground-state of the nuclear system, initial
conditions are defined to simulate excitations of giant resonances
in experiments with electromagnetic or hadron probes.
Frequencies of eigenmodes are determined from a Fourier
analysis of dynamical quantities. In this microscopic
model, self-consistent time-dependent 
mean-field calculations are performed for multipole excitations.
An advantage of the time-dependent approach is that no assumption
about the nature of a particular mode of vibrations has to be made.
Retardation effects for the meson fields are not
included in the model, i.e. the time derivatives
$\partial_t^2$ in the equations of motions for the meson
fields are neglected. This is justified by the large masses in the meson
propagators causing a short range of the corresponding
meson exchange forces. 
Negative energy contributions are included implicitly in
the time-dependent calculation, since the Dirac equation is
solved at each step in time for a different basis set~\cite{VBR.95}.
Negative energy components with respect to the original
ground-state basis are taken into account automatically,
even if at each time step the {\it no-sea} approximation is applied.

The description of nuclear
dynamics as a time-dependent initial-value problem is
intrinsically semi-classical, since there is no systematic
procedure to derive the initial conditions that
characterize the motion of a specific mode of the nuclear
system. The theory can be quantized by the
requirement that there exist time-periodic solutions of the
equations of motion, which give integer multiples of
Planck's constant for the classical action along one period
~\cite{RVP.96}.  For giant resonances the time-dependence
of collective dynamical quantities is actually not
periodic, since generally giant resonances are not
stationary states of the mean-field Hamiltonian. The
coupling of the mean-field to the particle continuum allows
for the decay of giant resonances by direct escape of
particles.  In the limit of small amplitude oscillations,
however, the energy obtained from the frequency of the
oscillation coincides with the excitation energy of the
collective state. In Refs.~\cite{VBR.95,RVP.96,Vre.97}
we have shown that the model reproduces 
experimental data on giant resonances in spherical nuclei.

\section{Monopole resonances and nuclear compressibility}

The study of isoscalar monopole resonances in nuclei
provides an important source of information on the nuclear
matter compression modulus $K_{\rm nm}$. This quantity is
crucial in the description of properties of
nuclei, supernovae explosions, neutron stars, and heavy-ion
collisions. In principle the value of $K_{\rm nm}$ can be
extracted from experimental energies of isoscalar
monopole vibrations in nuclei (giant monopole resonances GMR). 
However, the complete experimental data set on
isoscalar GMR does not limit
the range of $K_{\rm nm}$ to better than $200 - 350$ MeV.
Microscopic calculations of GMR excitation energies might
provide a more reliable approach to the determination of
the nuclear matter compression modulus.
Modern non-relativistic
Hartree-Fock plus RPA calculations, using both Skyrme and
Gogny effective interactions, indicate that the value of
$K_{\rm nm}$ should be in the range 210-220 MeV.
In relativistic calculations on the other hand, both
time-dependent and constrained RMF results indicate that
empirical GMR energies are best reproduced by an effective
force with $K_{\rm nm}\approx 250 - 270$ MeV.

In Ref.~\cite{Vre.97} we have performed time-dependent 
and constrained RMF calculations for monopole giant 
resonances for a number of spherical closed shell nuclei,
from $^{16}$O to the heavy nucleus $^{208}$Pb. For the effective
Lagrangian we have used six parameter sets, which differ
mostly by their prediction for $K_{\rm nm}$, but otherwise
reproduce reasonably well experimental data on nuclear properties.
The idea is to restrict the possible
values of the nuclear matter compression modulus, on the
basis of the excitation energies of giant monopole states
calculated with different effective interactions. In
addition to the four non-linear sets NL1, NL3, NL-SH and
NL2, we have also included two older linear parametrizations, 
HS and L1.
The sets NL1, NL-SH and NL2 have been extensively
used in the description of properties of finite nuclei
\cite{Rin.96}.  In order to bridge the gap between
NL1 ($K_{\rm nm} = 211.7$ MeV), and NL-SH ($K_{\rm nm} =
355.0$ MeV), we have also included a new effective
interaction NL3 ~\cite{LKR.97} ($K_{\rm nm} = 271.8$ MeV).
This new parameter set provides an excellent
description not only for the properties of stable nuclei,
but also for those far from the valley of beta stability.
From the energy spectra and transition densities calculated with
these effective forces, it has been possible to study the
connection between the incompressibility of nuclear matter
and the breathing mode energy of spherical nuclei.
For the isoscalar mode we have found an almost linear
relation between the excitation energy of the monopole
resonance and the nuclear matter compression modulus. 
For the determination of $K_{\rm nm}$
especially relevant are microscopic calculations of 
GMR excitation energies in heavy nuclei. The results of
TD RMF calculations for $^{208}$Pb are displayed in Fig. 1:
time-dependent monopole moments 
$\langle r^2(t)\rangle ~=~\frac {1}{A}\langle \Phi(t) |r^2 |\Phi(t)\rangle$
and the corresponding Fourier power spectra for the nonlinear
effective interactions.  As one would expect for a heavy
nucleus, there is very little spectral fragmentation and a
single mode dominates, at least for NL1 and NL3. The
experimental excitation energy $13.7\pm 0.3$ MeV is very
close to the frequency of oscillations obtained with the
NL3 parameter set: 14.1 MeV.  The calculated excitation
energy for the NL1 parameter set ($K_{\rm nm} =211.7$ MeV),
is approximately 1 MeV lower than the average experimental
value. For the linear effective forces HS and L1  the
oscillations are more anharmonic, and the monopole strength
is located well above the experimental GMR energy.

The effective interactions NL1 and NL3 seem to produce GMR
excitation energies which are quite close to the
experimental values. For these two parameter sets
we have calculated the isoscalar giant monopole
resonances in a number of closed-shell nuclei:
$^{40}$Ca, $^{56}$Ni, $^{100,114,132}$Sn, $^{90,122}$Zr, 
$^{146}$Gd. The results are shown in Fig. 2.
The energies of giant monopole states are determined from
the Fourier spectra of the time-dependent monopole moments,
and are displayed as function of the mass number. 
The NL1 excitation energies are systematically lower,
but otherwise the two effective interactions produce
very similar dependence on the mass number. The empirical
curve $E_x \approx 80~A^{-1/3}$ MeV is also included in the
figure, and it follows very closely the excitation energies
calculated with the NL3 parameter set. 
Similar results are obtained from constrained RMF calculations.
Both methods indicate that, in the framework of
relativistic mean field theory, the
nuclear matter compression modulus $K_{\rm nm} \approx 250
- 270$ MeV is in reasonable agreement with the available
data on spherical nuclei. This value is approximately 20\%
larger than the values deduced from recent
non-relativistic density dependent Hartree-Fock
calculations with Skyrme or Gogny forces.
It should be also emphasized that the excitation energy of the 
isovector monopole resonance in $^{208}$Pb, calculated with the 
NL3 effective force, is in excellent agreement with the 
experimental value $26 \pm 3$ MeV.

\section{Nonlinear dynamics of giant resonances}

Atomic nuclei provide excellent examples of quantum systems
in which the transition from regular to chaotic dynamics
can be studied. Signatures of chaotic
dynamics have been observed in correlations of nuclear
level distributions, and in the microscopic and collective
motion of the nuclear many-body system.
Especially interesting in this respect are giant resonances:
highly collective nuclear excitations whose properties,
excitation energies and widths, nevertheless reflect the
underlying microscopic dynamics. Theoretical studies have shown
that regular collective modes coexists with chaotic single-nucleon
motion: the adiabatic mean-field created by the nucleons averages
out the random components of their motion. However, it has also 
been shown that important differences exists between the 
isoscalar and isovector collective modes.

In particular, we have studied isoscalar and isovector
monopole oscillations in spherical nuclei \cite{Vre.97a}, but
analogous considerations apply to higher multipolarities.
In Fig. 3 results are shown of time-dependent
relativistic mean-field calculations for isoscalar
and isovector oscillations in $^{208}$Pb. Calculations have been
performed for the NL1 effective interaction.
In the isoscalar case both proton and
neutron densities are radially expanded, while for the
isovector mode the proton density is initially compressed
by the same amount. 
In Fig. 3a we plot the time series of the isoscalar
monopole moment $\langle r^2(t)\rangle$, and in Fig. 3b
the corresponding isovector moment
$< r^2_{\rm p}(t) > - < r^2_{\rm n} (t)>$ is displayed. The
isoscalar mode displays regular undamped oscillations,
while for the isovector mode strongly
damped anharmonic oscillations are observed.
On the right-hand panels we
plot the corresponding Fourier power spectra.
The Fourier spectrum of the isovector mode is strongly fragmented,
but the main peaks are found in the energy region $25 - 30$ MeV,
in agreement with the experimental data.
For the isoscalar mode, the time series of the monopole moment
and the Fourier spectrum show that the oscillations of the collective
coordinate are regular. On the other hand, the appearance of a
broad spectrum of frequencies seems
to indicate that the isovector oscillations are chaotic.
For time-series that result from linear physical processes the
Fourier analysis unfolds the characteristic frequencies which
are invariants of the dynamics, i.e. they classify the dynamics.
For nonlinear systems the corresponding analysis is somewhat
more complicated.

We have performed the analysis of phase spaces
reconstructed from time-series of collective dynamical
variables that characterize the isoscalar and
isovector oscillations. 
For the reconstruction of the dynamics two principal 
quantities have to be determined: the time delay
and the dimension of the phase space on which the attractor
unfolds. The delays which define the time-lagged variables 
have been determined from the first minima of the 
average mutual information function:
27 fm/c for the isoscalar, and 13 fm/c for the
isovector mode. For the embedding dimensions 
the method of false nearest neighbors has been used:
$d_E=3$ (isoscalar mode), and $d_E=4$ (isovector mode).
The reconstructed phase space can be represented by a
recurrence plot. For the
isoscalar mode the recurrence plot displays a pattern
representative for regular oscillations, 
for the isovector mode it indicates non-stationarity \cite{Vre.98}.
If the dynamics of a system is
deterministic, the ensemble of phase space trajectories
converges towards an invariant subset of the phase space -
the attractor. For chaotic dynamics the attractor has fractional
dimension, whereas the dimension is integer for regular dynamics. The
correlation dimension $D_2$ of the attractor can 
be numerically evaluated from
the log-log plot of the correlation integral $C_2(r)$ {\it vs}  
the distance $r$ in phase space,
for a set of increasing dimensions of the phase space.
As the embedding dimension increases, 
$D_2$ should saturate at a value equal to the
attractor's correlation dimension (Fig. 4). 
For the isoscalar mode, for $d\geq 3$, the correlation
dimension saturates at $D_2 = 2$; it does not saturate 
for the isovector mode, but slowly increases to some 
fractional value between 2 and 3. The fractional dimension
of the attractor would imply chaotic or stochastic dynamics.

The identification and quantification of the regular
or chaotic dynamics can be also based on the evaluation of
information-theoretic functionals. For the
time-dependent one-body nucleon densities, we have calculated
the von Neumann information entropy functionals.
The Fourier analysis has shown that the
entropy of the isoscalar mode contains the same information as
the dynamical variable, but the structure is more complicated
for the isovector mode. Giant resonances represent
collective oscillations of densities
with finite spatial extension, and therefore
provide excellent physical examples for the analysis of
systems that have spatial as well as temporal structure.
For a nonlinear system in chaotic regime, 
the influence of spatial motion on temporal chaos cab be studied:
what are the spatial correlations in a finite system that
displays chaotic oscillations of a collective dynamical variable?
These correlations have been described with a time-dependent
conditional entropy defined from a two-body nucleon density,
and the result is a high degree of two-body correlations for the
isoscalar mode, and very little spatial correlation
for the isovector oscillations of the nucleon density \cite{Vre.98}.
From the dynamical variables that
characterize the proton and neutron distributions, 
we have calculated the average mutual
information for the isoscalar and isovector modes (Fig. 5).
This function quantifies the information that is contained in
the dynamical variable of the neutron distribution,
about the proton subsystem, and vice versa.
The two curves that correspond to isoscalar and isovector
oscillations, are plotted as functions of the size of the box
in the linear embedding of the time-series. The average information
that is contained in the collective dynamical variable
of the proton density, about the neutron density, is more than
a factor three larger for the isoscalar mode.

Time-series, Fourier spectra, phase-space plots, 
Poincar\' e sections, autocorrelation functions, and
Lyapunov exponents \cite{Vre.97a} have also shown that
the motion of the collective coordinate is regular
for isoscalar oscillations, and that it becomes chaotic
when initial conditions correspond to the isovector
mode.

\newpage
\leftline{\Large {\bf Figure Captions}}
\parindent = 2 true cm
\begin{description}
\item[Fig.~1] Time-dependent isoscalar monopole moments
$<r^2>(t)$ and the corresponding Fourier power spectra for
$^{208}$Pb. The parameter sets are NL1, NL3, NL-SH and NL2.

\item[Fig.~2] Excitation energies of isoscalar giant
monopole resonances in spherical nuclei as
function of the mass number. The effective interactions
are: NL1 (squares) and NL3 (circles). The solid curve
corresponds to the empirical relation $\approx 80~A^{-1/3}$
MeV.

\item[Fig.~3] Time-dependent isoscalar and isovector monopole moments for
$^{208}$Pb.

\item[Fig.~4] Correlation dimension $D_2$ as function of
embedding dimension, for
isoscalar and isovector oscillations.

\item[Fig.~5] Mutual information between the time-dependent
mean square radii of the proton and neutron density distributions.
\end{description}
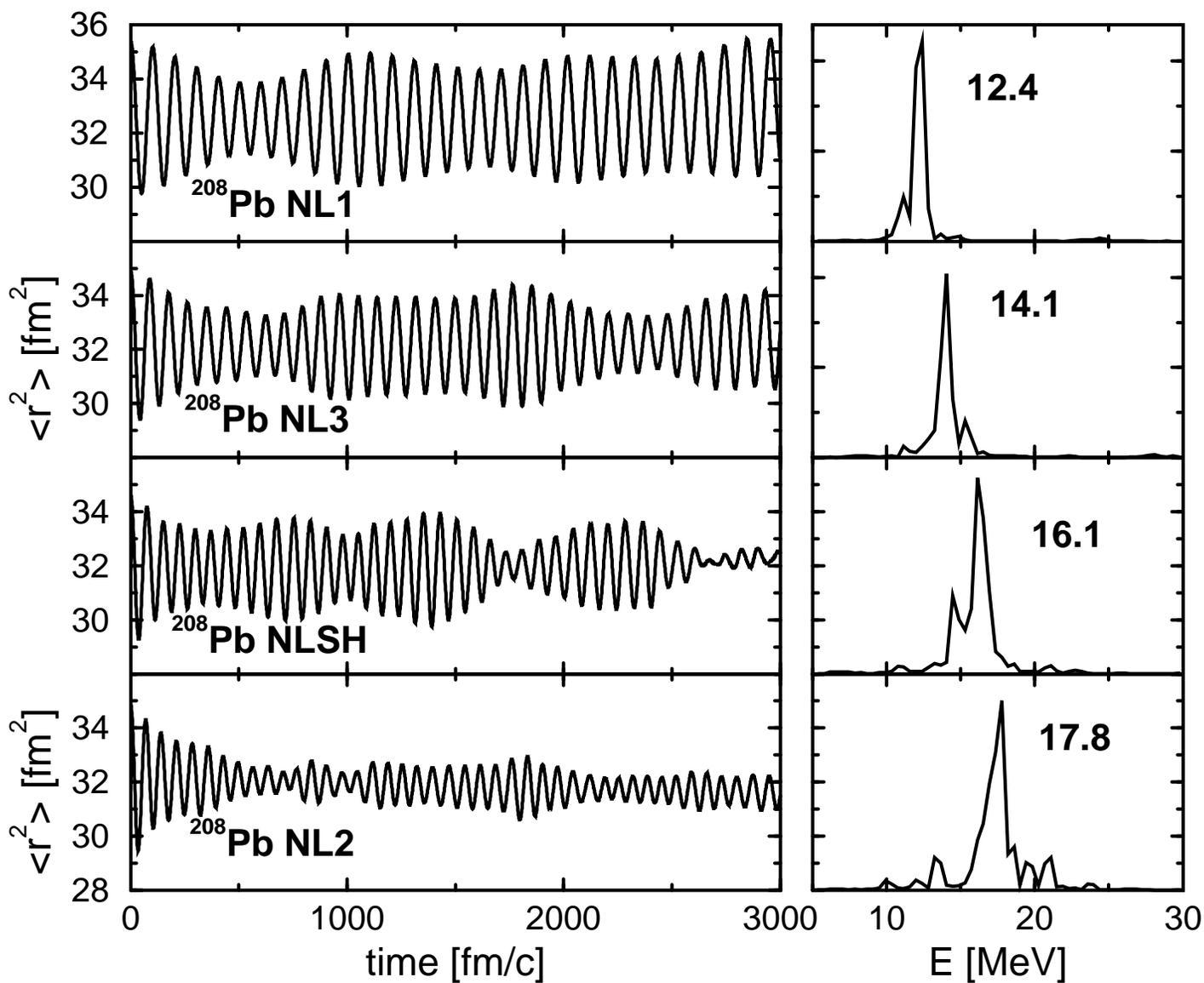
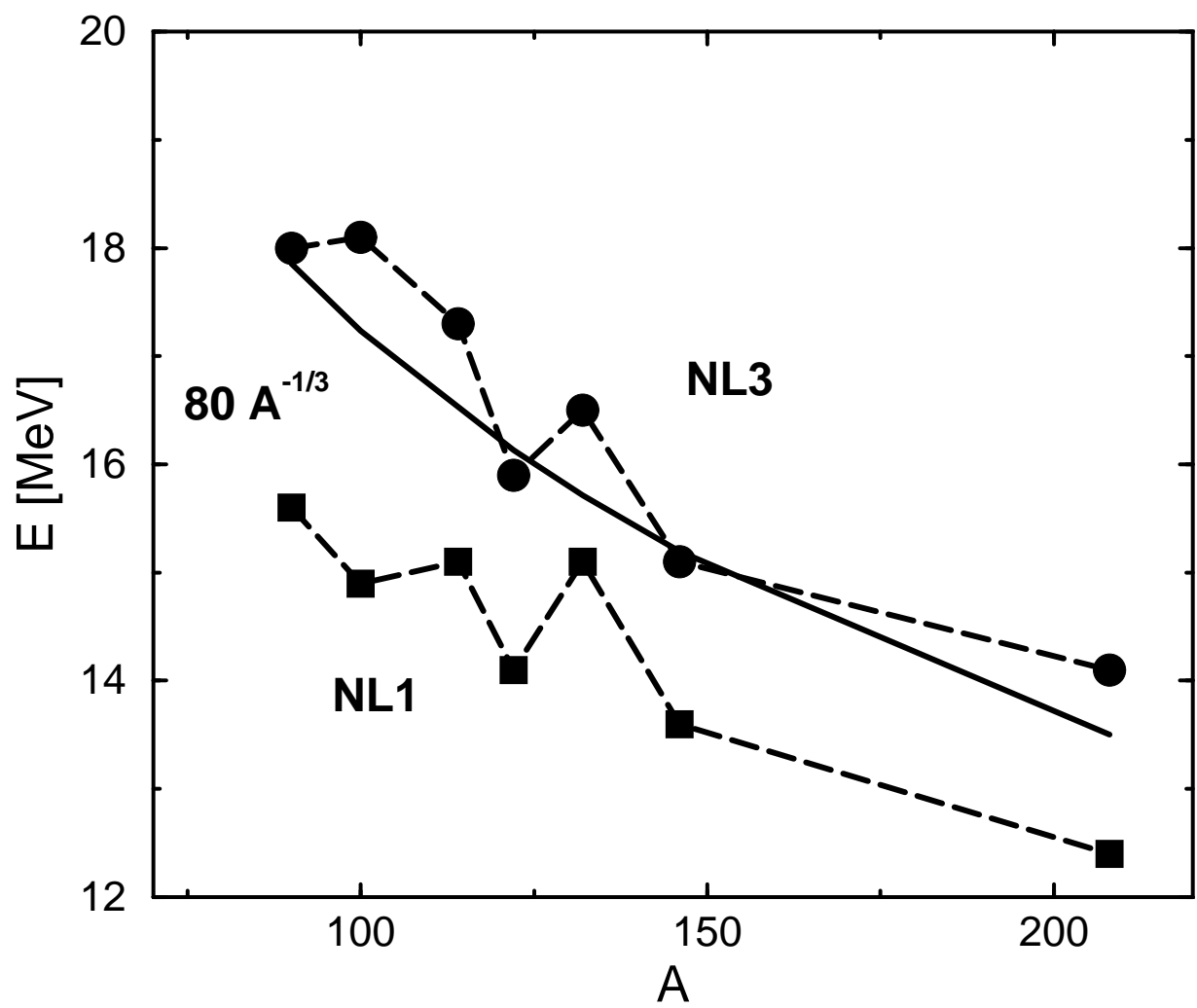
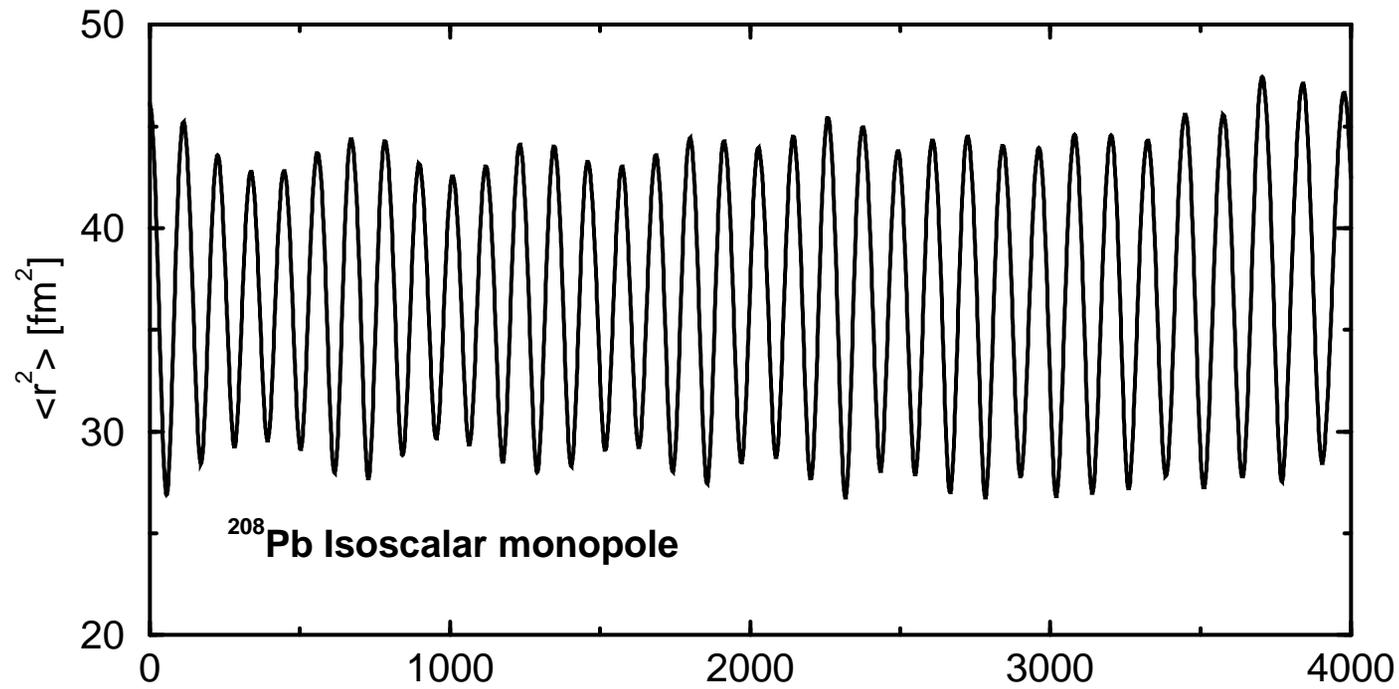
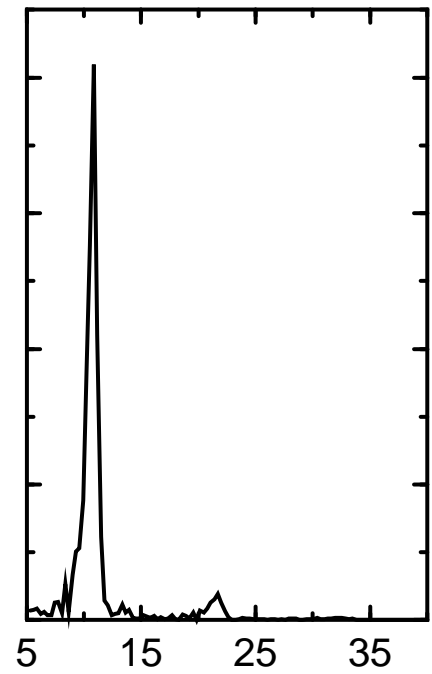
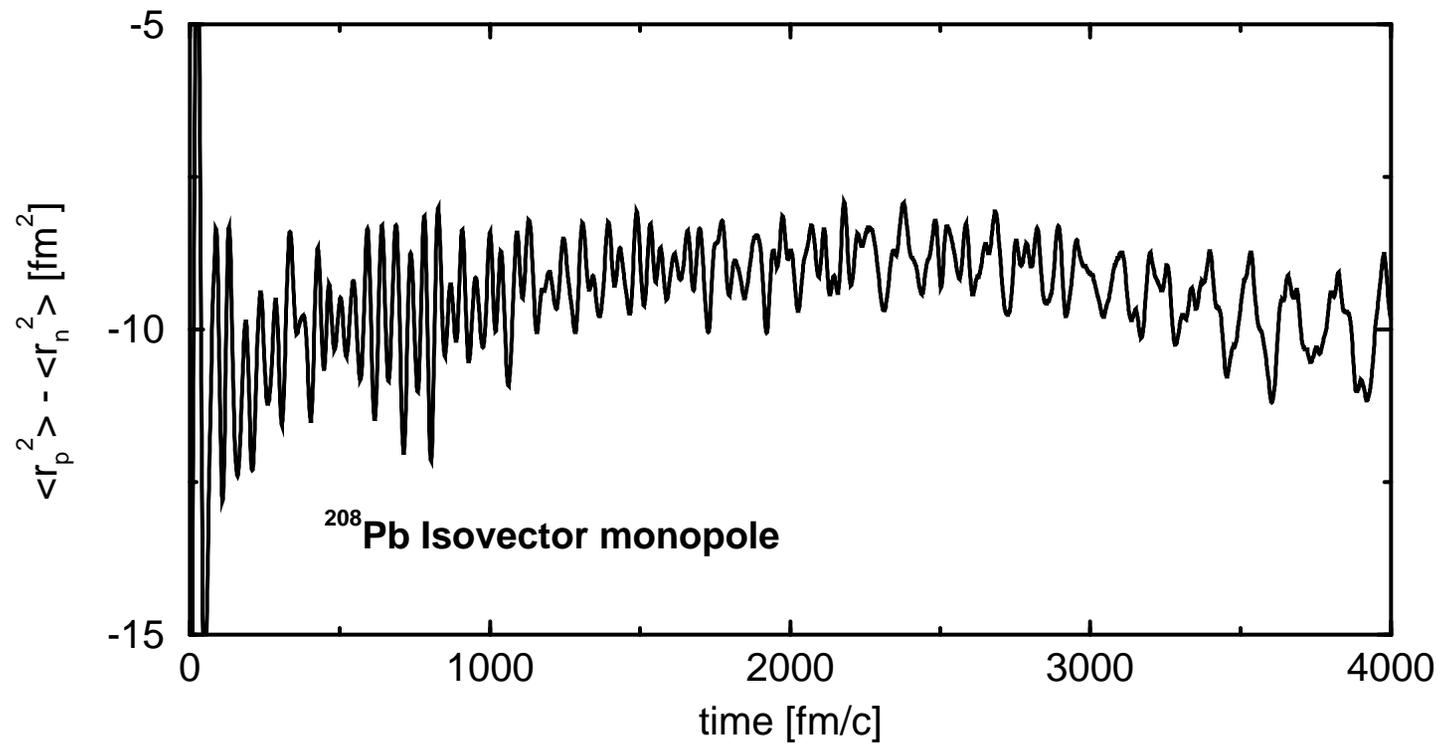
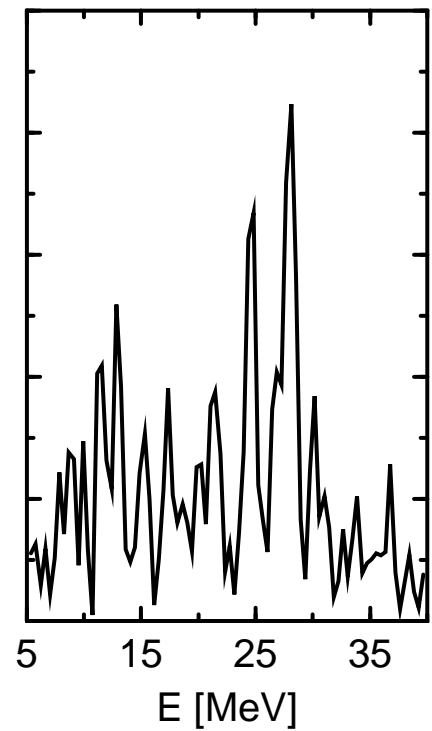
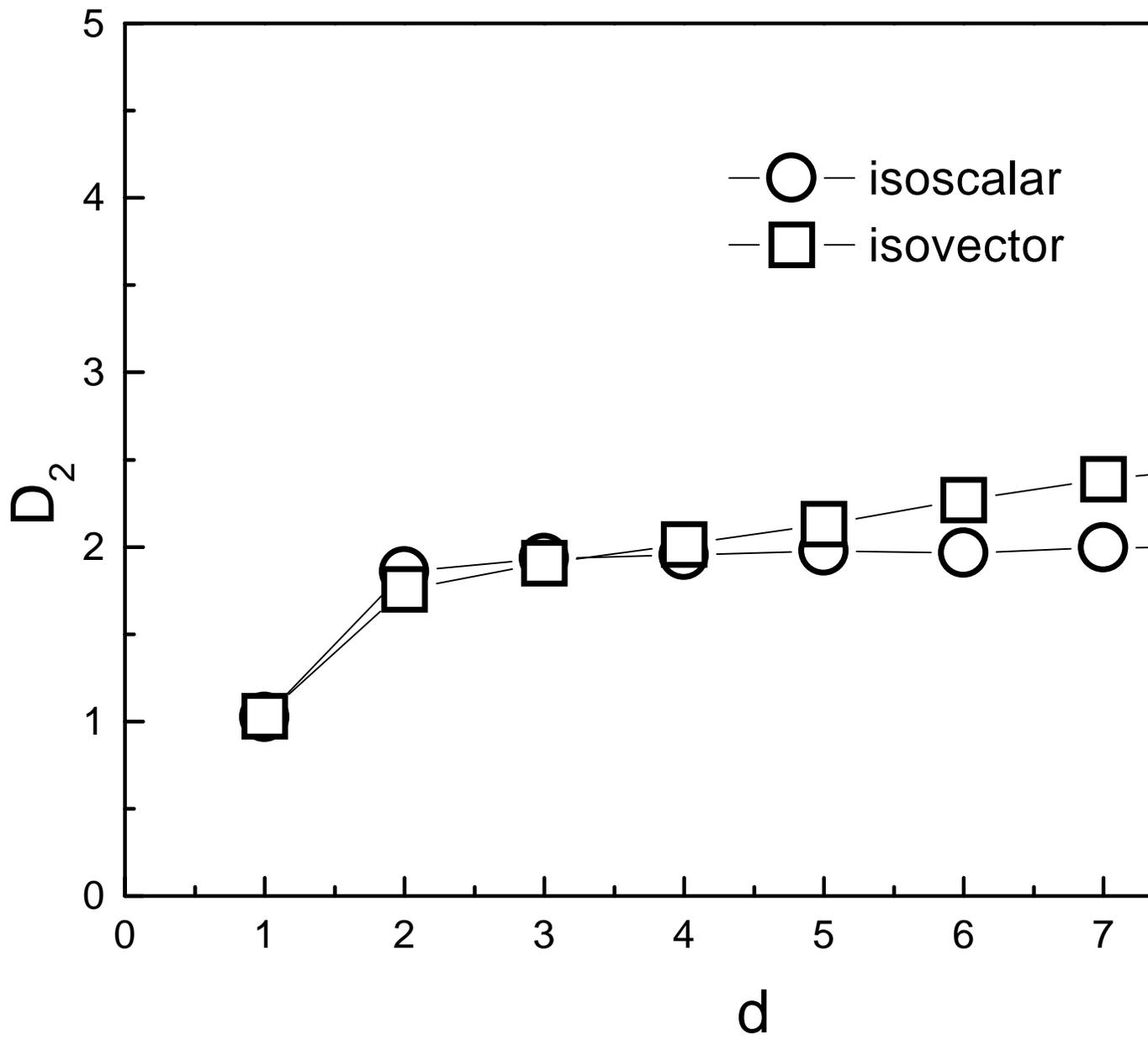
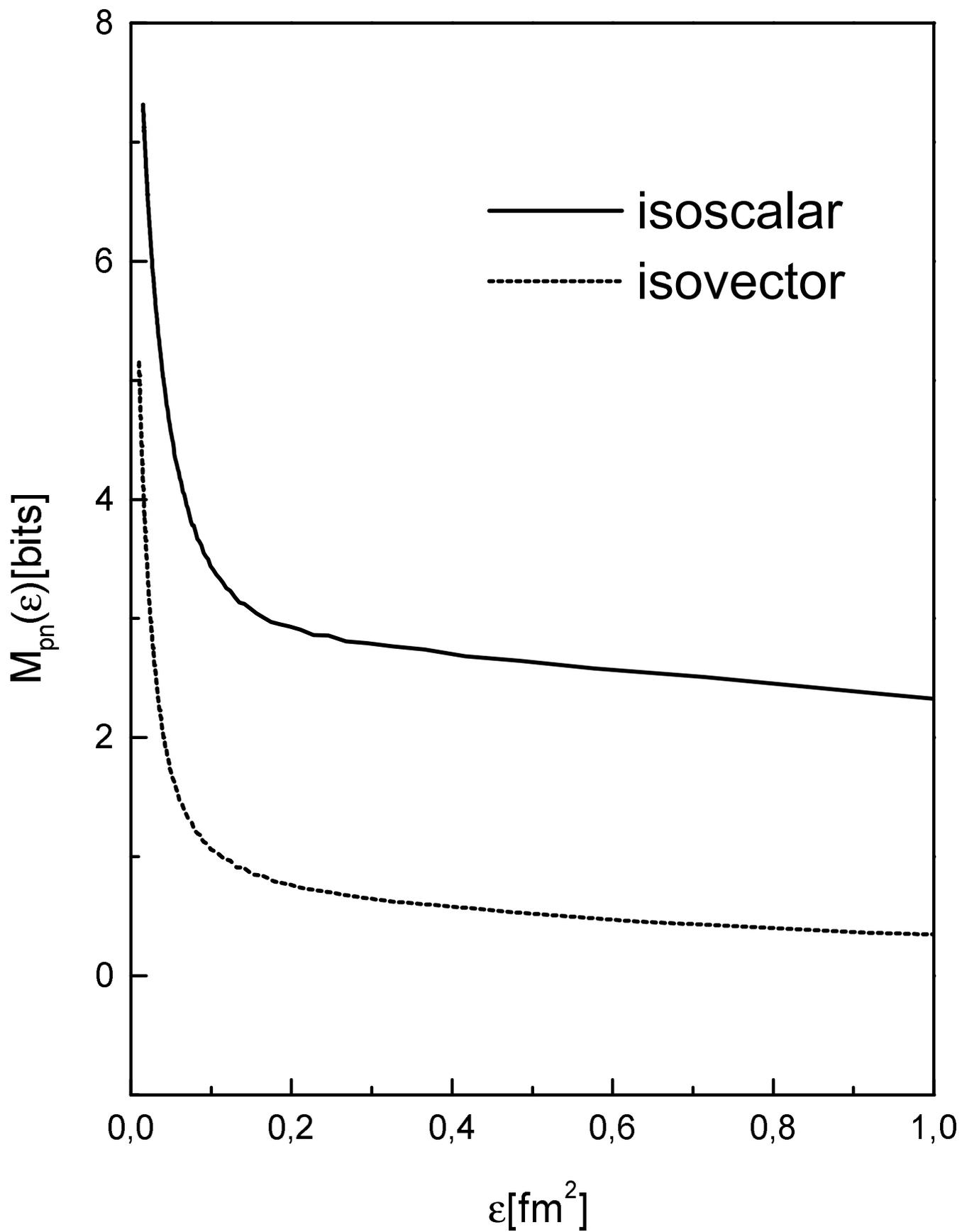
\end{document}